# Growth and engineering of perovskite SrIrO$_3$ thin films

Abhijit Biswas[a,b] and Yoon H Jeong[a,*]

[a] Department of Physics, Pohang University of Science and Technology, Pohang 790-784, South Korea

**Abstract**

5$d$ transition-metal-based oxides display emergent phenomena due to the competition between the relevant energy scales of the correlation, bandwidth, and most importantly, the strong spin-orbit coupling (SOC). Starting from the prediction of novel oxide topological insulators in bilayer ABO$_3$ (B = 5$d$ elements) thin-film grown along the (111) direction, 5$d$-based perovskites (Pv) form a new paradigm in the thin-film community. Here, we reviewed the scientific accomplishments in Pv-SrIrO$_3$ thin films, a popular candidate material for observing non-trivial topological phenomena. Although the predicted topological phenomena are unknown, the Pv-SrIrO$_3$ thin film shows many emergent properties due to the delicate interplay between its various degrees of freedom. These observations provide new physical insight and encourage further research on the design of new 5$d$-based heterostructures or superlattices for the observation of the hidden topological quantum phenomena in strong spin-orbit coupled oxides.

[b] **E-mail:** 01abhijit@gmail.com (Abhijit Biswas)

**\*Corresponding author:** yhj@postech.ac.kr (Yoon H Jeong)

# 1. INTRODUCTION

Strong spin-orbit-coupled 5$d$ transition metal oxides (TMOs), especially iridates (5$d$ Ir-based compounds) have become a new paradigm in scientific research with the discovery of a novel $J_{\text{eff}} = 1/2$ Mott insulator and, more intriguingly, for the observation of predicted topologically non-trivial phenomena [1-4]. Iridium is a heavy metal (atomic number, $z = 77$) and tends to exhibit strong spin-orbit coupling (SOC $\propto z^4$) of ~0.5 eV [5]. Due to the similar energy scales, in iridates, there is a complex interplay between the correlation ($U$), bandwidth ($W$), and SOC, making iridates extremely promising materials to explore new phenomena.

One family of iridates, the Ruddlesen-Proper series of $Sr_{n+1}Ir_nO_{3n+1}$ ($n = 1, 2,$ and $\infty$), show dimensionality-controlled systematic transition of material properties: for example, $Sr_2IrO_4$ ($n = 1$) is an antiferromagnetic $J_{\text{eff}} = 1/2$ Mott insulator, $Sr_3Ir_2O_7$ ($n = 2$) is barely an insulator, and endmember $SrIrO_3$ ($n = \infty$) is a paramagnetic semimetal [6,7]. It was demonstrated that because perovskite (Pv) $SrIrO_3$ shows lower conductivity than the expected highly metallic state for a 5$d$ transition metal with an extended $d$-orbital, the system is close to the metal-insulator phase boundary. This suggests that by applying external perturbations (strain, chemical pressure), the ground states of Pv-$SrIrO_3$ can be tuned, which may produce unconventional functionality.

About artificial designing of oxide heterostructures, one of the major breakthroughs in strong spin-orbit-coupled 5$d$ oxide research was the theoretical prediction of novel topological insulator phenomena by D. Xiao *et al*. It was predicted that with the atomically controlled bilayer growth of Pv-$SrIrO_3$ film along the (111) direction, due to the formation of honeycomb lattice geometry and strong SOC, two-dimensional topological insulator phenomena could be observed, with the characteristic signature of band inversion [8]. Later, it was shown that these heterostructures can form an antiferromagnetic ground state with trivial insulating nature, instead of a nontrivial topological state [9,10]. M. A. Zeb and H. Y. Kee predicted that Pv-$SrIrO_3$ would show a line of Dirac nodes near the Fermi level with small hole and electron pockets, due to the combination of the lattice structure and strong SOC [11-13]. They also proposed an interaction ($U$) vs. SOC ($\alpha$) phase diagram for Pv-$SrIrO_3$, showing that depending on the $U$ and $\alpha$ values, there might exist various ground states, e.g., nonmagnetic semimetal, magnetic metal, and magnetic insulator states. Recently, it was proposed that a novel topological crystalline metallic (TCM) state protected by crystalline lattice symmetry can also be achieved in Pv-$SrIrO_3$ by growth in a particular crystallographic direction [14].

In recent years, based on the above predictions, researchers have become highly interested in investigating this material. Although, initially it was observed that the stabilization of Pv-SrIrO$_3$ in high-quality thin film form was very challenging, but later many interesting studies have been published about this material. In this *review*, we summarize the key observations about the growth, properties, and engineering of Pv-SrIrO$_3$ heterostructures. We show that although topological insulator phenomenon is yet to be observed, the Pv-SrIrO$_3$ films show various emerging properties and provide new physical insight, which will be helpful for the future exploration of new functionalities in 5$d$ oxides.

## 2. CRYSTAL STRUCTURE OF SrIrO$_3$

At ambient pressure, SrIrO$_3$ forms a monoclinic structure ($a$ = 5.604 Å, $b$ = 9.618 Å, $c$ = 14.17 Å, $\beta$ = 93.26°) of space group *C2/c*. The crystal structure was first determined by J. M. Longo *et al* [15]. At high pressure ($P$ = 40 kbar) and temperature ($T$ = 1000 °C), it transforms into a distorted orthorhombic structure ($a$ = 5.60 Å, $b$ = 5.58 Å, $c$ = 7.89 Å) of space group *Pbnm* (**Fig. 1**). Because SrIrO$_3$ adopts cubic close packing with high density at high pressure, the structure transforms into an orthorhombic structure with an approximately 3% reduction in volume [15]. The pseudo-cubic unit cell of the distorted orthorhombic structure yields $a_{pc} \approx a/\sqrt{2} \approx b/\sqrt{2} \approx c/2 \approx 3.96$ Å.

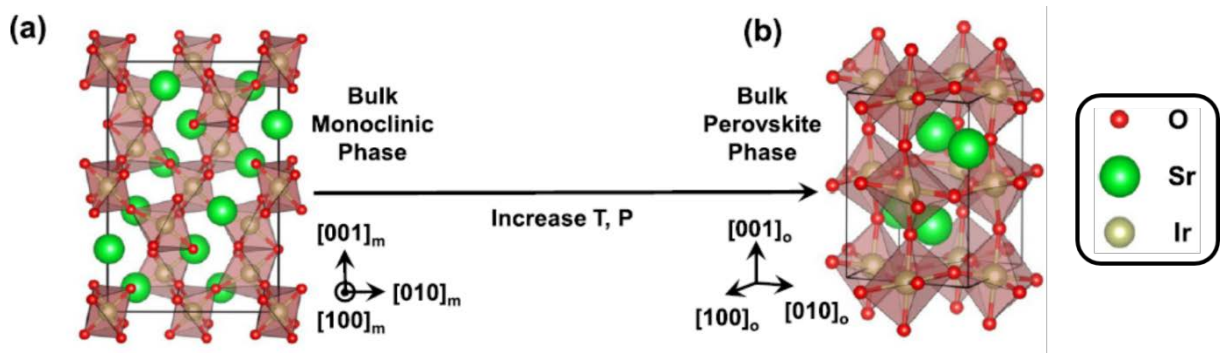

**Fig. 1.** (Color Online) (a) The ambient-pressure monoclinic phase turns into a (b) distorted orthorhombic structure at high pressure ($P$) and temperature ($T$). The pseudo-cubic lattice constant for the orthorhombic phase is $a_{pc} \approx 3.96$ Å. Reprinted with permission from T. J. Anderson *et al.*, Appl. Phys. Lett. **108**, 151604 (2016). Copyright 2016 AIP Publishing LLC [55].

Several groups have reported the synthesis of the polycrystalline form of Pv-SrIrO$_3$ [15-20], and have refined the crystal structure using x-ray and neutron diffraction methods to obtain insight into the crystal structure. The individual IrO$_6$ octahedra in Pv-SrIrO$_3$ are relatively rigid. Two independent octahedral tilts ($\chi$, $\phi$) are characteristic of a *Pbnm*-type orthorhombic structure, where $\chi$ is the out-of-phase tilt along the (110) direction and $\phi$ is the in-phase tilt along the (001) direction. For Pv-SrIrO$_3$, $\chi$ = 11.5° and $\phi$ = 8.7° at *T* = 300 K, and the values are independent of temperature at *T* ≤ 300 K **[18]**. The Ir-O bond length (~2.016 Å at *T* = 300 K) is equal along all three directions, and the average Ir-O-Ir bond angle $\langle\psi\rangle$ is ~154.1° **[19]**. Even at *T* = 1100 K, the material shows orthorhombic structure. As polycrystals contain lots of grains and suffer from high porosity, growing single crystal is highly desirable. However, due to its meta-stable nature, growing single crystal of Pv-SrIrO$_3$ is technically difficult as single crystal of Pv-SrIrO$_3$ is only stable at high *P,* thus making ambient studies and applications unfeasible **[21]**. By taking the advantage of thin film growth method this problem can be circumvented, and one can stabilize high-quality Pv-SrIrO$_3$ films by providing the effective *P* exerted from the underlying substrate. Films can be grown from its ultrathin limit to its superlattice form. Since the emerging phenomena were predicted for the orthorhombic phase of SrIrO$_3$, we mainly focus on the study of Pv-SrIrO$_3$ films.

## 3. THIN FILM RESEARCH ON PEROVSKITE SrIrO$_3$

### I. GROWTH OF EPITAXIAL SrIrO$_3$ THIN FILMS

Epitaxial thin films of Pv-SrIrO$_3$ are generally grown by: (1) pulsed laser deposition (PLD), (2) molecular beam epitaxy (MBE), (3) metal organic chemical vapor deposition (MOCVD) and (4) RF (radio frequency) magnetron sputtering **[22-43]**. For thin film deposition, one can use either a homemade polycrystalline SrIrO$_3$ target or a commercially available target. The polycrystalline SrIrO$_3$ target is prepared by conventional solid-state reaction. Because Pv-SrIrO$_3$ is a meta-stable phase, during growth, the oxygen partial pressure is generally kept at high pressure (e.g., *P* = 10-300 mTorr). Although it is conventional to use a high growth temperature (*T*$_G$) for most thin film growth due the kinetics, but for the volatile nature of Ir at high temperature, it is better to grow Pv-SrIrO$_3$ films at lower growth temperature (e.g., *T*$_G$ < 700 °C) **[34]**. For better crystallinity and to compensate for the loss of oxygen, the films are post-annealed and cooled down slowly.

Several key observations were made during the deposition of thin films. During target ablation, the flume size decreases significantly after several hundred pulses/site **[25]**. During ablation, the target becomes Ir-rich and possibly forms an IrO$_2$ metallic surface, which reflects the incident laser, forming sparks that come out from the target. To minimize these effects, the target needed to be polished after each deposition. Transmission electron microscopy (TEM) and other structural analyses show that films grown up to ~40 nm maintain the coherency of the underlying substrate (**Fig. 2(a)**). Films thicker than 40 nm show polycrystalline quality as the peak width broadens (**Fig. 2(b)**), showing two microstructures composed of both meta-stable orthorhombic and stable monoclinic SrIrO$_3$ phases **[28]**. Therefore, the maximum thickness that limits the growth of high-quality single-crystal Pv-SrIrO$_3$ thin films is ~40 nm. Epitaxial Pv-SrIrO$_3$ films are also sensitive to air, and their electrical properties degrade while exposing the film to ambient conditions and after lithographic processing **[40]**. Therefore, an alternative solution is to use a capping layer for precise measurements for future device fabrication.

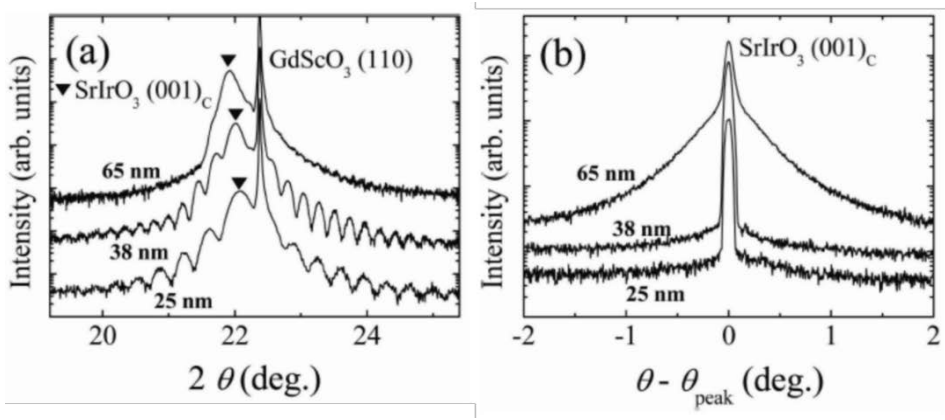

**Fig. 2.** (Color Online) (a) X-ray diffraction of Pv-SrIrO$_3$ films grown on the best lattice-matched GdScO$_3$ (110) substrate. Although the thickness fringes are observed for 25 nm and 38 nm films, thicker film (65 nm) loses the crystallinity, with a very broad peak. (b) A broad full width half maxima (FWHM) is observed for film thicker than 38 nm. Reprinted with permission from S. Y. Jang *et al.*, J. Korean. Phys. Soc. 56, 1814 (2010). Copyright 2010 Korean Physical Society **[25]**.

## II. TRANSPORT PROPERTIES IN BULK SrIrO$_3$

Before addressing the thin film properties, we consider the properties of polycrystalline Pv-SrIrO$_3$. The bulk polycrystalline Pv-SrIrO$_3$ shows metallic behavior with resistivity ($\rho$) ~1.5 m$\Omega\cdot$cm at $T$ = 300 K (**Fig. 3(a)**) **[18]**. The resistivity follows Fermi-liquid behavior at low temperature ($\rho \propto T^2$), with the characteristic electron-electron scattering at low $T$. The residual resistivity ratio ($R_{300K}/R_{2K}$) is very small (< 2), which corresponds to semimetal-like characteristics. Positive $B^2$-like ($B$ = magnetic field) magnetoresistance (MR) is observed;

typical for a metal. Magnetic measurements demonstrate that Pv-SrIrO$_3$ is Pauli paramagnetic [16,18]. Overall, Pv-SrIrO$_3$ is a correlated paramagnetic semi-metal.

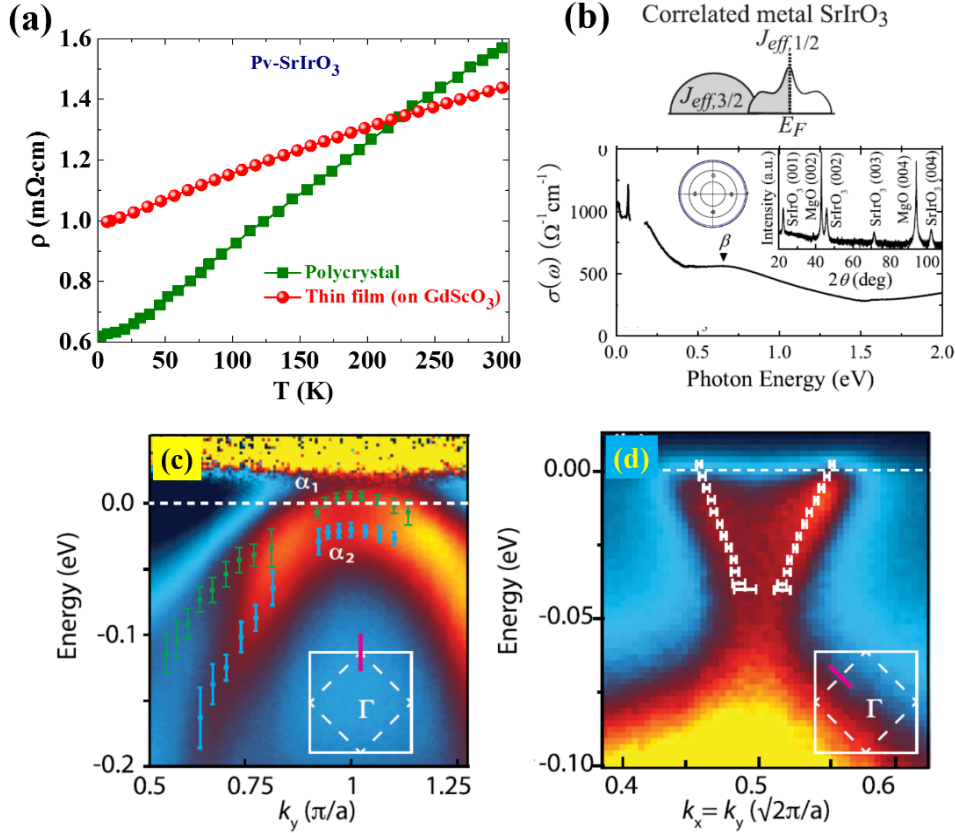

**Fig. 3.** (Color Online) (a) Resistivity of Pv-SrIrO$_3$ in its bulk polycrystalline as well as in thin-film form (on lattice-matched GdScO$_3$ (110) substrate). The resistivity at room is almost similar for both phases, with a very low residual resistivity ratio (RRR). In both phases, low-temperature Fermi-liquid ($\rho \propto T^2$) behavior is observed. (b) Optical conductivity $\sigma(\omega)$ spectra of SrIrO$_3$ shows metallic behavior. $\beta$ corresponds to the transition from the $J_{eff} = 3/2$ bands to the upper Hubbard band (UHB). The ARPES measurement shows (c) two heavy hole-like bands at $(0, \pi)$ and (d) a light, almost linearly dispersive electron-like band at $(-\pi/2, \pi/2)$. The polycrystalline resistivity was adapted with permission from P. E. R. Blanchard *et al.*, Phys. Rev. B **89**, 214106 (2014). Figs. (b), (c) and (d) were reprinted with permission from S. J. Moon et al., Phys. Rev. Lett. 101, 226402 (2008); Y. F. Nie *et al.*, Phys. Rev. Lett. 114, 016401 (2015). Copyright 2015 American Physical Society **[6,18,44]**.

Several groups have been successful in growing SrIrO$_3$ thin films, especially in the perovskite form, on various substrates **[22-43]**. The first successful growth of Pv-SrIrO$_3$ thin film was reported by Y. K. Kim *et al.* by MOCVD **[24]**. Because the pseudo-cubic lattice constant of Pv-SrIrO$_3$ is $a_{pc} \approx 3.96$ Å, it perfectly matches the lattice constant of the GdScO$_3$

(110) substrate, producing the most natural state film, without imposing tensile or compressive strain. The most natural state film of thickness 35 nm (~88 unit cells) is fully metallic, with a room temperature resistivity of ~1.5 mΩ·cm (**Fig. 3(a)**) [32], similar to the polycrystalline sample. The carrier concentration and Hall mobility's are ~$10^{-19}$ cm$^{-3}$ and ~100-200 cm$^2$V$^{-1}$S$^{-1}$ at low *T*, which is characteristic of a semimetal [35]. The positive MR at *T* = 5 K for the thin film follows a $B^2$-like power law, as expected for a normal paramagnetic metal. Optical conductivity measurement shows the Drude-like response for the electrons in the $J_{\text{eff}} = 1/2$ bands in the energy region below 0.5 eV, and an inter-band transition from the $J_{\text{eff}} = 3/2$ to $J_{\text{eff}} = 1/2$ state at 0.7 eV (**Fig. 3(b)**)[6,33], demonstrating the effect of spin-orbit coupling on the $J_{\text{eff}}$ states in the case of Pv-SrIrO$_3$, identifying it as a special type of oxide.

The dimensionality-controlled metallicity in Pv-SrIrO$_3$ is a well-accepted characteristic, i.e., with an increasing number of IrO$_2$ layers, the coordination number, and thus the bandwidth, increases [6]. However, recent direct observation of the band picture by in-situ measurement of surface-sensitive angle-resolved photoemission spectroscopy (ARPES) of Pv-SrIrO$_3$ film shows a completely different picture. The bandwidths are surprisingly narrower than the Sr$_2$IrO$_4$ (~0.5 eV)[6], and are on the order of ~0.3 eV [38, 44]. Additionally, there is the unusual coexistence of heavy hole-like and light electron-like bands (**Fig. 3(c), (d)**), which indicates the possible mixing of $J_{\text{eff}} = 1/2$ and $J_{\text{eff}} = 3/2$ states. By using ARPES and synchrotron x-ray analysis, it was also shown that mirror symmetry-protected Dirac line nodes near the Fermi level (as predicted theoretically) in Pv-SrIrO$_3$ thin films can actually be lifted by breaking the Pbnm-symmetry of the crystal structure, highlighting the important role in crystal structure in observing topological phenomena [38,42].

The above observations suggest that there is a complex interplay between the crystal geometry, SOC, dimensionality, and correlation in bulk Pv-SrIrO$_3$ thin films. The engineering of the Pv-SrIrO$_3$ phase thus would be an interesting subject to observe the emerging phenomena.

### III. THICKNESS VARIATION: BULK TO ULTRATHIN LIMIT

Changing the thickness of a film, i.e., reducing the dimensionality from its bulk form to the ultrathin limit, is an important parameter in inducing novel functionalities in thin-film physics. Reducing the film thickness not only changes the film's properties, but ultrathin films also show novel quantum phenomena. Reducing a film's thickness generates several key physical aspects: (a) it reduces the coordination number and thus reduces the band filling and bandwidth, which increases the effective correlation (*U/W*) and makes the system an insulator, (b) it increases the grain size, grain boundaries, and step edges, which increases the effective disorder, thus making the system an insulator, and (c) it increases the strain and associated complex chemistry between the substrate and the film, producing novel phenomena.

For Pv-SrIrO$_3$ thin films, several groups have reported the observation of a metal-insulator transition (MIT) when reducing the film thickness [27,32,37]. Starting with fully metallic bulk film, when lowering the film thickness (~3-4 nm), the system makes a transition from semi-metal to an insulating state (**Fig. 4(a)**). The origin of this insulating state is due to the effect of disorder (the grain size and grain boundaries increase). Disorder driven two-dimensional weak-localization (2D-WL) phenomena and associated negative MR are observed for ultrathin films (**Fig. 4(b), (c)**). The 2D-WL fitting shows that the carriers are mainly scattered by electron-phonon interactions [27]. Therefore, the observed MIT when varying the thickness is due to *Anderson localization* associated with the fully localized carrier.

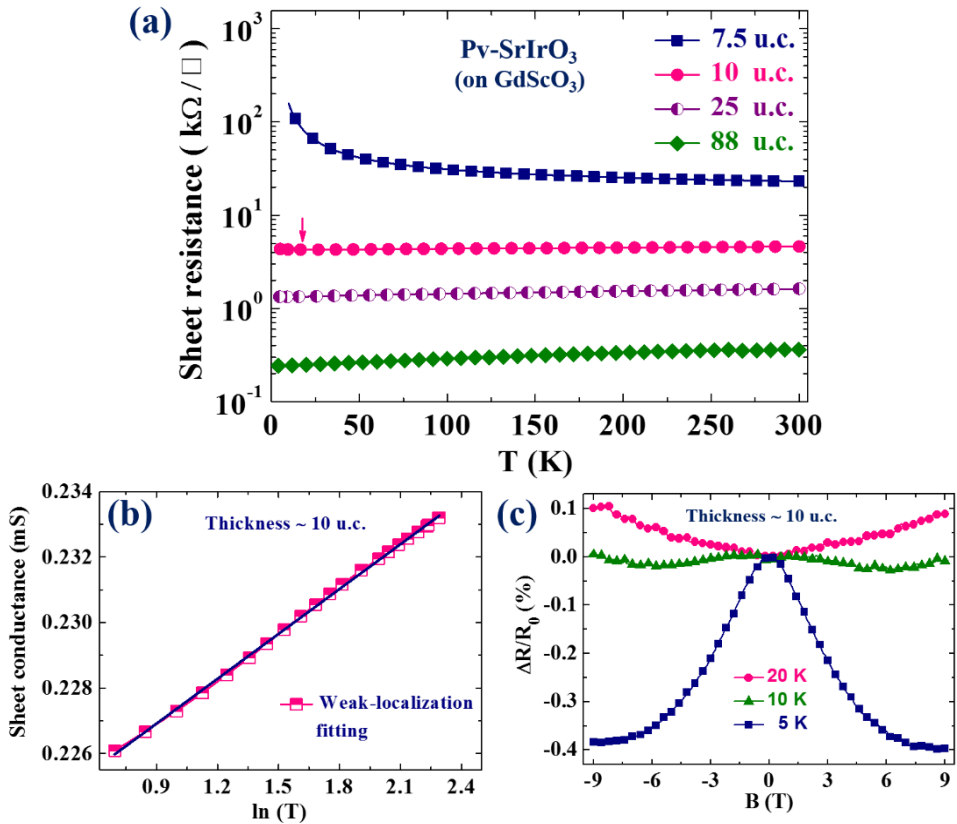

**Fig. 4.** (Color online) (a) Metal insulator transition of Pv-SrIrO$_3$ grown on lattice-matched GdScO$_3$ (110) substrate while reducing the film thickness. (b) Two-dimensional weak-localization fitting ($\sigma \propto \ln T$) of the increase in resistivity (pink arrow in Fig. (a)). (c) Observed negative magnetoresistance in the weak-localization regime. Reprinted with permission from A. Biswas *et al.*, J. Appl. Phys. **116**, 213704 (2014). Copyright 2014 AIP Publishing LLC [32].

## IV. STRAIN ENGINEERING: CHANGING THE LATTICE MISMATCH

Exerting strain on a film using different lattice mismatch substrates is one pioneering way to engineer functionality in metal oxide thin films. Strain in a Pv-ABO$_3$ film tilts the octahedra from their pristine positions, which changes the hopping path of the conduction electrons. This effectively changes the bandwidth ($W$) of the system by following the relationship

$$W \propto \frac{cos\Theta}{d^{3.5}}$$

where $d$ is the B-O bond length and $\Theta = (\pi - \psi)/2$ is the buckling deviation of the B-O-B bond angle $\langle\psi\rangle$ from $\pi$.

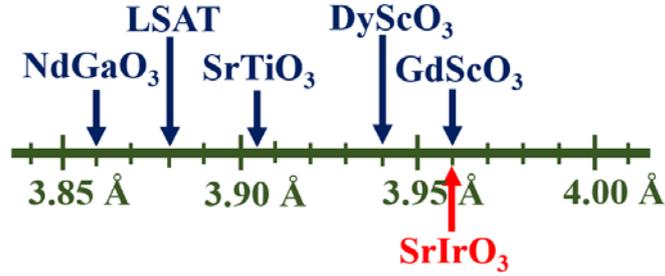

**Fig. 5.** (Color Online) Cubic and/or pseudo-cubic lattice constants of Pv-SrIrO$_3$ and the metal oxide substrates used to induce strain, mainly the compressive strain.

Several groups have studied strain engineering in Pv-SrIrO$_3$ thin films, and all groups came to the conclusion that the MIT can be achieved by exerting compressive strain on the film **[30,32,35,36]**. The compressive strain can be induced by using lower lattice constant substrates, e.g., DyScO$_3$ (110), SrTiO$_3$ (001), LSAT (001), NdGaO$_3$ (110), and LaAlO$_3$ (001) (**Fig. 5**). The systematic increase in the compressive strain indicates a systematic increase of resistivity. At high compressive strain, the system transitions to a fully insulating state (**Fig. 6**). L. Zhang *et al.* observed the nonlinear to linear Hall effect transition and the decrease of carrier concentration, with the compensation of electron hole balance by increasing the compressive strain **[35]**. Because the Ir-O bond is rigid, the increase in compressive strain decreases the Ir-O-Ir bond angle, which reduces the electron hopping between the 5$d$ Ir orbitals and shifts the Fermi level, thus inducing an insulating state. Additionally, although the system is three-dimensional, but at lower temperature, the increase in resistivity in the three-dimensional films follows the two-dimensional WL model. The magnetoresistance in the weak-localization regime also shows a positive correction in resistivity **[32]**. The optical conductivity measurements demonstrate that even fully insulating films show a Drude-like response without opening an optical gap **[30]**, indicating a semi-metallic state of the film with complex interplay between its various degrees of freedom, which has also been supported by various theoretical calculations **[45,46]**. Also for compressive-strained films, at low magnetic field and low *T* negative MR was observed, signifying the role of weak localization and strong SOC **[27,30]**.

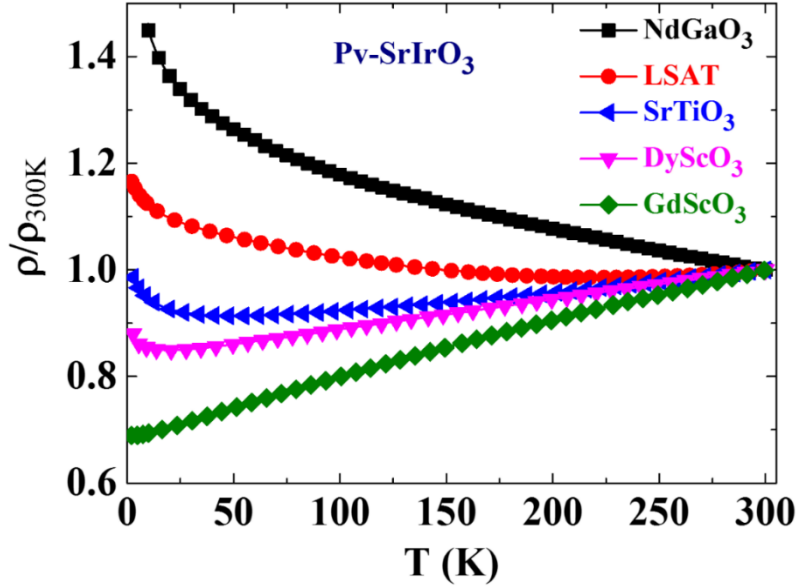

**Fig. 6.** (Color Online) Compressive strain-induced metal-to-insulator transition in Pv-SrIrO$_3$ thin film. The film on the lattice-matched GdScO$_3$ (110) substrate shows fully metallic behavior, but fully insulating behavior is observed for the film on the NdGaO$_3$ (110) substrate, which produces high compressive strain. Intermediate strained samples on DyScO$_3$ (110) and SrTiO$_3$ (001) substrates also show metallic behaviors with upturn in resistivity at low *T*. The resistivity data were adapted with permission from J. H. Gruenewald *et al*., J. Mater. Res. 29, 2491 (2014); A. Biswas *et al*., J. Appl. Phys. **116**, 213704 (2014). Copyright 2014 Cambridge University Press. Copyright 2014 AIP Publishing LLC **[30,32]**.

During the transition from the metallic to insulating state, the resistivity power law ($\rho \propto T^\varepsilon$) for all the metallic films changes with the strain energy. With the increase in compressive strain, the power law in resistivity changes from $\varepsilon = 4/5$ to 1 to 3/2 **[32]**. This is surprising because in the presence of spin fluctuation and disorder, quantum critical systems usually show a change in the resistivity power law **[47]**. Since Pv-SrIrO$_3$ is a paramagnetic metal **[16,18]**, the interplay between SOC, correlation, and disorder is probably the origin of the variation of the power law in resistivity. However, one cannot neglect the importance of localized magnetic moments **[39,48]**, which remains to be clarified by future microscopic measurements.

The above observations in the strain-dependent MIT scenario suggest that there is an interplay between the crystal geometry, correlation, and disorder in the presence of strong SOC, which is fully responsible for not only obtaining MIT but also for various contradictory characteristics. The observed MIT in the compressive-strain-induced scenario is due to *Mott-Anderson-Griffiths* type of localization associated with correlation and disorder in the presence of strong SOC.

## V. ENGINEERING OF ARTIFICIAL HETEROSTRUCTURES

Atomically controlled bilayer 5$d$ TMOs grown along the (111) direction of a perovskite substrate form a buckled honeycomb lattice (**Fig. 7**). Due to the geometry effect and strong SOC, this process can result in non-trivial topological insulator phenomena **[8]**. From a material design point of view, it requires atomic scale layer-by-layer control of the film. It is not straightforward and is rather challenging to obtain specifically designed bi-layer superlattices because the (111) direction of Pv-SrIrO$_3$ consists of alternating charged layers of SrO$_3^{4-}$ and Ir$^{4+}$. The diverging surface energy makes it energetically unfavorable to produce high-quality thin films with unit-cell, atomically flat, step-terrace structures.

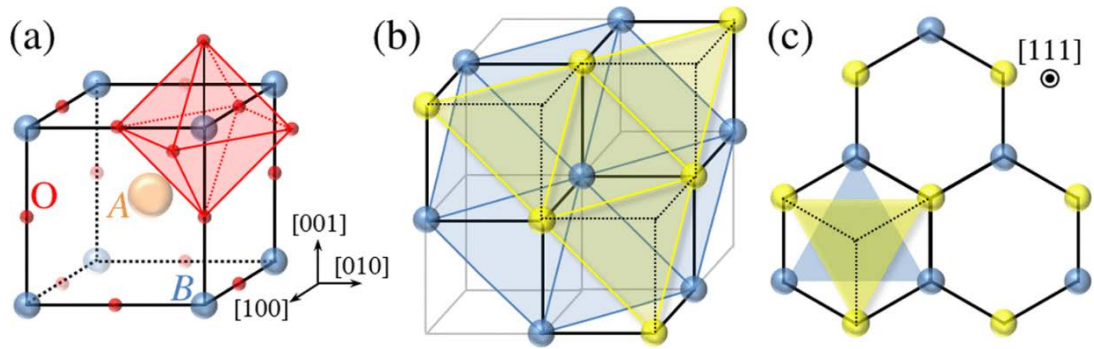

**Fig. 7.** (Color Online) (a)-(c) The Pv-ABO$_3$ structure forms a buckled honeycomb structure with two (111) planes (yellow and blue). Reprinted with permission from D. Hirai *et al.*, APL Mater. **3**, 041508 (2015). Copyright 2015 AIP Publishing LLC **[52]**.

J. Matsuno *et al*. first successfully synthesized superlattices of [(SrIrO$_3$)$_m$/SrTiO$_3$] along the (001) direction and observed that by reducing the number of SrIrO$_3$ layers ($m$), the system transforms from a semi-metal to a magnetic insulator state (**Fig. 8(a)**) **[49]**. Magnetic ordering was observed only in insulating films with $m \leq 3$. The ordering temperatures were $T \sim 100$ K for $m \leq 2$, and $T \sim 140$ K for $m \leq 1$ (**Fig. 8(b)**). The magnetism is canted antiferromagnetic type (moment ~0.02 $\mu_B$/Ir for $m = 1$) and is only along the film's in-plane direction. This is a consequence of the Dzyaloshinskii-Moriya (DM) interaction associated with the in-plane rotation of the IrO$_6$ octahedra due to the strain from the underlying substrate **[49-51]**. Their consistent transport and magnetic data implies couple of important things: 1) ultrathin SrIrO$_3$ can be magnetic, and 2) magnetism, rather than disorder is the driving force in MIT.

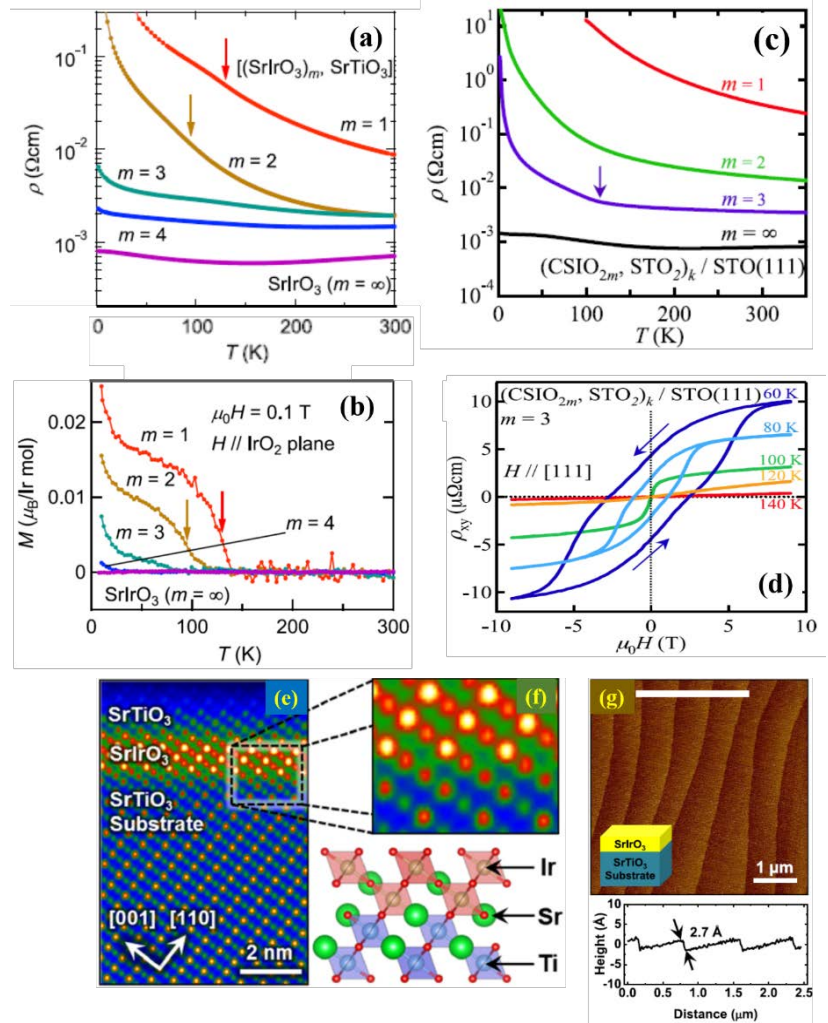

**Fig. 8.** (Color Online) (a)-(b) Resistivity and magnetization of [(SrIrO$_3$)$_m$,SrTiO$_3$] (001) superlattices. (c)-(d) Resistivity and Hall effect of [(Ca$_{0.5}$Sr$_{0.5}$IrO$_3$)$_{2m}$,(SrTiO$_3$)$_2$]$_k$/SrTiO$_3$ (111) superlattices with $m$ = 1, 2, 3, 4 and ∞. As the layer thickness decreases, the superlattice properties change from semi-metal to magnetic insulator. Down arrows indicate the magnetic ordering temperature. (e)-(f) Atomically resolved high-angle annular dark-field (HAADF)-scanning transmission electron microscopy (STEM) image shows the epitaxial arrangement of three-bilayer Pv-SrIrO$_3$ grown along the (111) direction of a SrTiO$_3$ substrate, with five SrTiO$_3$ capping layers to prevent surface contamination. (g) Atomically flat surface of high-quality Pv-SrIrO$_3$ film showing a single-bilayer step-terrace structure. Reprinted with permission from J. Matsuno *et al.*, Phys. Rev. Lett. **114**, 247209 (2015); D. Hirai *et al.*, APL Mater. **3**, 041508 (2015); T. J. Anderson *et al.*, Appl. Phys. Lett. **108**, 151604 (2016). Copyright 2014 AIP Publishing LLC, Copyright 2016 American Physical Society **[49, 52, 55]**.

Later, D. Hirai *et al.* realized that the strain due to the large lattice mismatch between the film and substrate ($a_{pc}$ ~ 3.96 Å for Pv-SrIrO$_3$, and $a_{pc}$ ~ 3.905 Å for SrTiO$_3$) adds more difficulty for growing layer-by-layer controlled films **[52]**. They reduced the lattice constant of the film by replacing half of the Sr with Ca. This takes the system more closer to SrTiO$_3$ because the isostructural Pv-CaIrO$_3$ pseudo-cubic lattice constant is $a_{pc}$ ~ 3.86 Å, and fortunately the bulk properties of Pv-SrIrO$_3$ and Pv-CaIrO$_3$ thin films are almost same **[53,54]**. They successfully grew atomically controlled [(Ca$_{0.5}$Sr$_{0.5}$IrO$_3$)$_{2m}$/(SrTiO$_3$)$_2$]$_k$ superlattices on SrTiO$_3$ (111) substrate. These superlattices also show magnetic insulator phenomena with decreasing iridates layer thickness (*m*), due to the delicate interplay between the electron correlation and strong SOC (**Fig. 8(c), (d)**).

Recently, T. J. Anderson *et al.* successfully stabilized bilayer of pure Pv-SrIrO$_3$ film on SrTiO$_3$ (111) substrate and confirmed the formation of a honeycomb lattice structure, which is the model system for observing the predicted novel topological phenomena (**Fig. 8(e)-(g)**) **[55]**. Deeper investigation of the properties of these heterostructures, especially observation of the band topology by ARPES, should be given priority in the future.

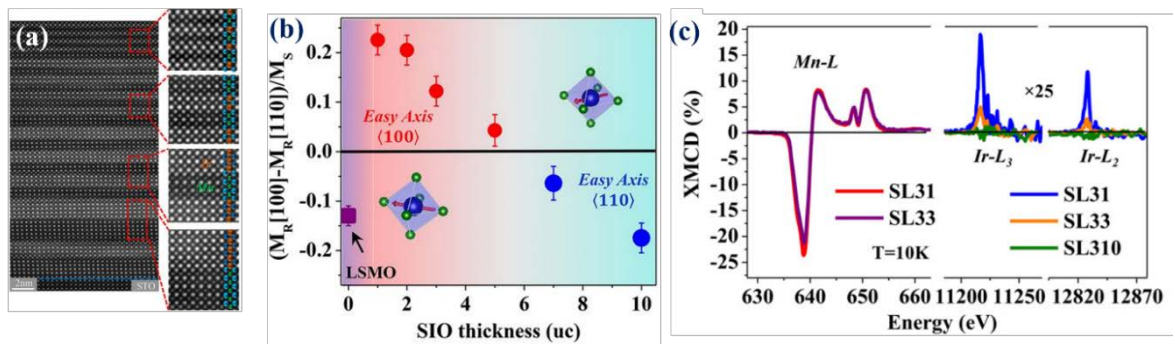

**Fig. 9.** (Color Online) (a) Atomically resolved high-angle annular dark-field (HAADF)-scanning transmission electron microscopy (STEM) image of (La$_{2/3}$Sr$_{1/3}$MnO$_3$)$_3$/(SrIrO$_3$)$_m$ (*m* = 1 to 10) superlattices grown on a (001) SrTiO$_3$ substrate. (b) The rotation of the magnetic easy axis of La$_{2/3}$Sr$_{1/3}$MnO$_3$ from (110) to (100) depends on the SrIrO$_3$ layer thickness. (c) The X-ray magnetic circular dichroism (XMCD) measurement shows the presence of both an Ir-L$_3$ and Ir-L$_2$ edge. Reprinted with permission from D. Yi *et al.*, Proc. Natl. Acad. Sci. **113**, 6397 (2016). Copyright 2016 National Academy of Sciences, USA **[56]**.

Designing artificial heterostructures by combining iridates with other well-studied transition metal oxides to produce novel phenomena is also important for future electronic applications. To take the advantage of the strong SOC and the paramagnetic nature of Pv-SrIrO$_3$, researchers have synthesized artificial superlattices by combining a ferromagnetic metal La$_{2/3}$Sr$_{1/3}$MnO$_3$ with a paramagnetic metal Pv-SrIrO$_3$ (**Fig. 9(a)**) **[56]**. When changing the thickness of Pv-SrIrO$_3$ within a few unit cells, the magnetic easy axis of La$_{2/3}$Sr$_{1/3}$MnO$_3$ rotates between the (110) and (001) crystallographic directions due to the formation of a novel SOC state (mixing of $J_{eff} = 1/2$ and $J_{eff} = 3/2$ states; a different scenario from the only $J_{eff} = 1/2$ state in Sr$_2$IrO$_4$ as observed based on the atomic selection rule **[7]**) with a relatively large orbital-dominated Ir moment (**Fig. 9(b),(c)**). These observations are particularly important for designing new artificial heterostructures with atomic-scale engineering for device applications.

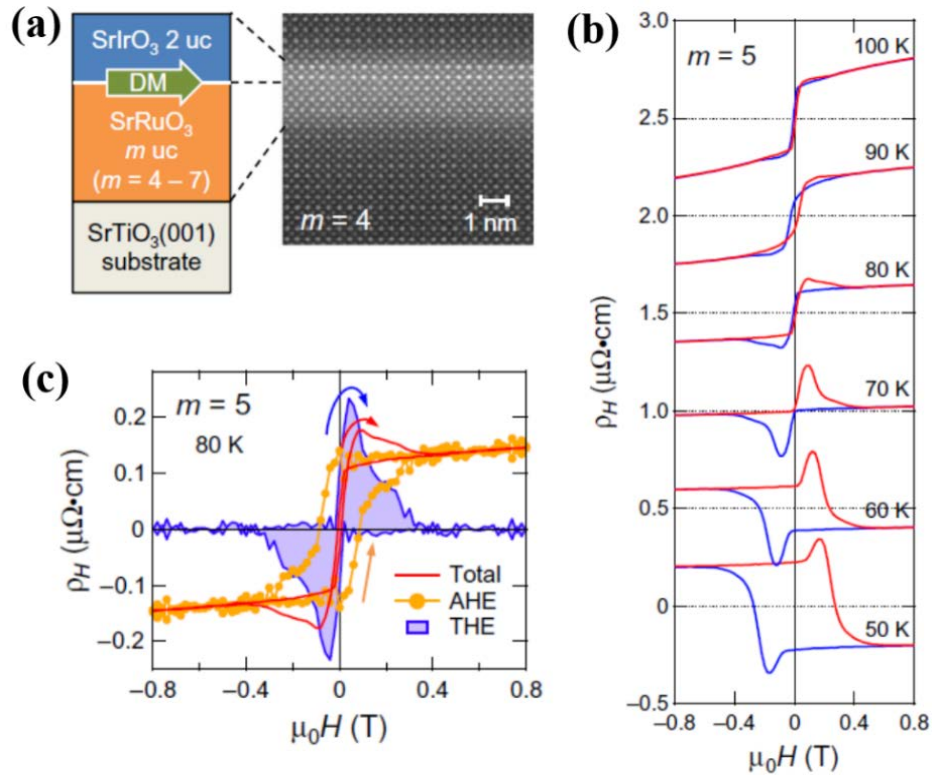

**Fig. 10.** (Color Online) (a) Atomically resolved high-angle annular dark-field (HAADF)-scanning transmission electron microscopy (STEM) image of the (SrIrO$_3$)$_2$/(SrRuO$_3$)$_m$ superlattice, where $m$ = 4-7. (b) An anomaly in the Hall resistivity (hump) was observed between 0.8 and 2.1 Tesla. (c) Extracted contribution from the anomalous Hall effect (AHE) and topological Hall effect (THE). Reprinted with permission from J. Matsuno *et al.*, Sci. Adv. **2**, e1600304 (2016). Copyright 2016 American Association for the Advancement of Science **[57]**.

Recently, J. Matsuno *et al.* observed a novel topological Hall effect (THE) in SrRuO$_3$-SrIrO$_3$ superlattices **[57]**. THE is different from ordinary Hall effect (consequence of Lorentz force) or anomalous Hall effect (consequence of magnetization in ferromagnets). Sometimes the Hall effect can originate from the scaler spin chirality which is attributed with the Berry phase in real space, thus can be termed as THE. THE is a special phenomenon in condensed matter physics as it hosts magnetic skyrmions, which are basically topologically protected spin swirling texture. In forming magnetic skyrmions, the DM interaction play a significant role. As the DM interaction arises from the combination of broken inversion symmetry and SOC, they grew superlattices of bilayer $SrIrO_3$ with varying layer thickness of ferromagnetic $SrRuO_3$, i.e., $(SrIrO_3)_2/(SrRuO_3)_m$, where $m$ = 4-7 (**Fig. 10(a)**). Although the resistivity and magnetization features are quite similar to $SrRuO_3$, but they observed that at lower $SrRuO_3$ thickness, there is a clear anomaly (a hump) in the Hall resistivity hysteresis (**Fig. 10(b)**), which is attributed to THE (**Fig. 10(c)**), where the interface DM interaction plays a significant role. Theoretically they have also shown that in this THE regions, skyrmions are energetically stable. Although the real space observation of this phenomena need more experiments, they provide a new platform for the design of artificial superlattices to observe new topological effects.

## 4. CONCLUSIONS AND FUTURE PROSPECTS

Pv-$SrIrO_3$ falls into a special class of material that has drawn extensive research interest. It contains a heavy metal on the *B*-site (5*d* Ir) with strong SOC, producing a variety of novel phenomena. Due to the technical difficulties, perovskite single crystals have not been successfully grown, and researchers have only focused on the polycrystalline and/or thin-film form. The thin-film growth shows peculiar characteristics that are different from the growth of other transition metal oxides. The meta-stable Pv-$SrIrO_3$ films, from a transport point of view, show metal-insulator transitions, not only when reducing the film thickness but also when inducing compressive strain on the film. The origin of these MITs are different. One is associated with the Anderson localization (thickness variation), and the other one is due to Mott-Anderson localization (strain-dependent). Unusual reduction in band-width (w.r.t $Sr_2IrO_4$) was observed directly by ARPES measurements with the formation of both electron-like and hole-like bands, showing more intrinsic complex picture of interplay between correlation, crystal geometry, and SOC in Pv-$SrIrO_3$. A possible scenario of novel spin-orbital state, i.e. the mixing of $J_{eff}$ = 1/2 and 3/2 state has also been proposed. More interesting phenomena for iridates were observed when grown in the superlattice form ($SrIrO_3/SrTiO_3$ or $Ca_{0.5}Sr_{0.5}IrO_3/SrTiO_3$). The system makes the transition from a paramagnetic semimetal to an antiferromagnetic insulating state depending on the superlattice thickness. Researchers have also grown $La_{2/3}Sr_{1/3}MnO_3/SrIrO_3$ superlattices and found that the magnetic easy axis of $La_{2/3}Sr_{1/3}MnO_3$ can be altered by changing the $SrIrO_3$ layer thickness, which would be useful for spintronic purposes. More recently, THE was observed in $SrRuO_3/SrIrO_3$ superlattices while lowering the SrRuO$_3$ thickness, a step towards observing topological phenomena in 5$d$ oxides. From application point of view, Pv-SrIrO$_3$ also show high electro-catalytic activity **[58]**.

Although a variety of phenomena has been observed for Pv-SrIrO$_3$, many more remain to be explored in the future. The current research has been focused on high-quality thin film growth and the transport phenomena in Pv-SrIrO$_3$ films or its superlattices. Changing the material functionality by doping (especially with 3$d$, 4$d$ elements) is an interesting issue, as it will not only generate chemical pressure to stabilize lower energetic crystal structures, but also will change the energy strength of the effective correlation and SOC **[59,60]**. Due to its very compatible crystal structure, making highly metallic films (conductivity close to copper) by doping (e.g. La$^{3+}$) or creating oxygen vacancies would make iridates an additional candidate for electrode for thin films. Inducing magnetism in the bulk Pv-SrIrO$_3$ film could provide additional spin degree of freedom to come into effect to tune the functionality. Due to its structural distortions, ultrathin Pv-SrIrO$_3$ film may show anisotropic MR, which would be useful for device applications. As several groups have succeeded in growing layer-controlled films, so future spectroscopic characterization will be important for microscopic insight, especially to deeply investigate the topological characteristics suggested by theoretical calculations. Additionally, although researchers have already started growing new artificial iridate superlattices, but to expand the field of research in 5$d$-based artificial oxides, other materials especially A$_2$BIrO$_6$ double perovskites should be grown to observe novel phenomena **[11, 61-64]**.

As observed, after overcoming various difficulties in growing these layer-by-layer controlled films, researchers have synthesized bilayer Pv-SrIrO$_3$ (111) films; a model system for obtaining non-trivial topological phenomena in oxides. Since 5$d$ TMOs are an important class of materials for obtaining the emergent phenomena, in the future, one should consider observing the band topology not only in iridates but also in other systems containing strong spin-orbit-coupled elements (e.g., W, Re, Os) **[65]**. Finally, one should consider the possible applications for manufacturing devices that use the advantage of strong SOC in the 5$d$ transition metal oxides. We hope this timely overview of results in Pv-SrIrO$_3$ thin films will be helpful for researchers working in this field and will pave the way for future explorations of topological quantum phenomena in 5$d$ transition-metal-based oxide heterostructures.

ACKNOWLDGEMENTS

YHJ was supported by National Research Foundation (NRF) of Korea (SRC-2011-0030786 and 2015R1D1A1A02062239).